\documentclass[journal=jacsat,manuscript=article,dvipsnames]{achemso}
\usepackage{graphicx}
\usepackage{amsmath}
\usepackage{color}

\title{Surface Termination and Band Alignment in 2D Heterostructures}
\author{Raheel Hammad*, Snehith Adabala and Soumya Ghosh*}
\date{}

\begin{document}

\maketitle

\def\sgc#1{\textcolor{red}{#1}}
\def\rhc#1{\textcolor{blue}{#1}}

\begin{abstract}
Heterostructures are ubiquitous in many optoelectronic devices and as photocatalysts. One of the key features of a heterojunction is the proper band alignment between the two materials. Estimation of the correct relative band positions with density functional theory (DFT) based electronic structure calculations is often constrained by the accuracy and cost associated with the various DFT functionals. In this study, we introduce a novel computational approach that achieves band alignments closely matching experimental results with the widely used PBE functional. We specifically examine the well-documented MoO$_3$/MoS$_2$ system, a type-II heterojunction. In our setup, the MoS$_2$ layers are kept as it is but for MoO$_3$ the individual layers are chosen differently. These alternative layers have higher surface energy, and hence, the band edges are higher than the conventional layers. This shift in band edges of the alternative MoO$_3$ layers changes the band alignment in MoO$_3$/MoS$_2$ heterojunction from type-III to the experimentally observed type-II character. We also extend this computational strategy to additional systems, demonstrating its versatility and effectiveness.

\end{abstract}

\textbf{Keywords:} Type-II heterostructure, valence band offset, density functional theory, band gap

\section{Introduction}
Heterostructures have shown a lot of potential for their use in rechargeable batteries, supercapacitors, solar cells, and photocatalysis.\cite{xiong2020strain,kumar2023photo,gao2022dual,britnell2013strong,prominski2022porosity,li2015state,li2013preparation}
%The semiconducting heterostructures can be classified into three types based on their band alignments, namely straddling (type I), staggered (type II), and broken-gap (type III). In type I heterojunctions both the Valence band maxima (VBM) and Conduction band minima (CBM) of one material lies in between the VBM and CBM of other material. Hence, under photoexcitation these heterojunctions accumulate the electrons and holes in the material with lower gap. In Type III heterojunctions  VBM of one material has higher potential than CBM of the other one. 
In type II(staggered) heterojunction, the conduction band minimum (CBM) of one material lies in between the valence band maximum (VBM) and the CBM of the other. 
If the material with higher CBM is photoexcited to generate electron-hole pairs then the excited electrons are transferred to its partner but the holes stays in the same material, resulting in a spatial separation of electron-hole pairs, thus reducing their recombination rate. Hence these staggered heterojunctions are crucial in optoelectronic applications and photocatalysis. \cite{zhong2020type,zhou20182d,kumar2023photo,li2015state,li2013preparation}\\

Due to various desirable electronic and mechanical properties 2D transition metal dichalcogenides are favored in several of these applications.\cite{manzeli20172d} The experimental band positions of MoS$_2$ and MoO$_3$ and their band gaps suggest that the combination can form a type II heterojunction.\cite{zhou2024moo3,saadati2021single} Due to the presence of direct band gap, high surface area, good stability and high mobility, MoS$_2$ have been employed in various applications in conjunction with MoO$_3$.\cite{li2016high,zhou2024moo3,li2024high,saadati2021single}
Multiple experimental studies have reported the formation of type II heterojunction between MoS$_2$ sheets and either bulk or layers of MoO$_3$.\cite{zhou2024moo3,chen2020high,hao2021comparative,zhou2014vertically,li2024high}. 
However, several computational studies fail to reproduce this trend in band alignment. \cite{shahrokhi20212d,gao2019degenerately, kc2016electronic,zhang2022activating}. In these computational studies, the van der Waal heterostructure of MoO$_3$ and MoS$_2$ have been shown to form a type III heterojunction. In an interesting study, Shahroki et. al. have shown that type II character can be recovered for an explicit heterojunction if the oxygen atoms of MoO$_3$ layer, which are in direct contact with MoS$_2$, are replaced with sulphur atoms.\cite{shahrokhi20212d} \\

In our study, we develop a computational methodology to generate a type-II band alignment between MoS$_2$ and MoO$_3$ within the framework of density functional theory (DFT).\cite{martin2020electronic} This methodology employs an alternate layer cleavage scheme for MoO$_3$ without altering their chemical composition. This alternative layering has a higher surface energy that results in band shifts with respect to the MoS$_2$ bands, thus forming a type II heterojunction.
This computational scheme is generally applicable to any 2D semiconductor materials. To showcase our point we investigate a few other material combinations that have been experimentally recognized as type-II, e.g.MoO$_3$/MoS$_2$, MoTe$_2$/MoO$_3$ and CdS/CuSbS$_2$. We believe that this computational scheme would be highly useful for studying chemical processes involving 2D semiconductors where the type-II band alignment is essential to capture the important electronic effects.
% ====================================================================================================
\section{Methods}
The DFT calculations employed the Generalized Gradient Approximation (GGA) based PBE functional \cite{perdew1996generalized} as well as the hybrid functional HSE06 \cite{krukau2006influence} for benchmark purposes. These calculations were performed in the Vienna ab-initio simulation package (VASP) \cite{kresse1993ab,kresse1996efficiency}  using a plane wave basis set with the projector augmented plane-wave method (PAW) \cite{kresse1999ultrasoft}. A kinetic energy cutoff of 500eV for used for MoO$_3$/MoS$_2$ and MoS$_2$/MoTe$_2$, while a cutoff of 280 was used for CuSbS$_2$/CdS. The energy and force convergence criteria were set to be \(10^{-4}\) eV and 0.02 eV/\AA respectively for all the calculations. The Gamma Centred Monkhorst scheme of grid density 0.025 (\(2 \pi\) /Å\(^{-1}\)) is used in each direction to sample the Brillouin zone. As pointed out by Yan Hua Lei et.al \cite{lei2012dft+}, accurate MoO$_3$ calculations require Hubbard correction on top of DFT. Therefore we have used DFT+U calculations introduced by Dudarev et. al. \cite{dudarev1998electron} with a Hubbard parameter of 5.0 eV for a system with MoO$_3$. Moreover, in accordance to other literature studies, a Hubbard parameter of 5.0 eV for Cu and 4.0 for Cd were used for CuSbS$_2$ \cite{de2017characterization} and CdS \cite{khan2020investigating}, respectively. For all multilayered systems, DFT+U was combined with the Grimes et.al dispersion correction (D3) to account for the long-range interaction \cite{grimme2010consistent}.

For computing the band alignment of isolated materials, we have referred the energy levels to the vacuum level by computing planar averaged Hartree potential. The VBM of bi-layer MoS$_2$(2L) is computed with HSE06 functional relative to its macroscopic average, which in turn is referenced to the vacuum level in a slab calculation. For bulk MoO$_3$ calculations (with HSE06), the VBM is referenced to the vacuum level using the following protocol. We first compute the electronic structure of bulk MoO$_3$ and multi-layered slabs of MoO$_3$ [010](Figure S1) in conjunction with vacuum. The lateral dimensions of the slab and the bulk supercells were kept the same while the geometry was relaxed in both cases.  
Central layers in the slab are supposed to represent the bulk and using the macroscopic average of this region one can reference the VBM of the bulk to the vacuum \cite{fu2020band,conesa2021computing}. We then computed the valence band offset (VBO) between slabs of MoS$_2$ and bulk MoO$_3$.
The simplified formula for the above method is given below (see SI for details). 
\begin{align}
\mathrm{VBO} & = E^\mathrm{VBM}_{l-\mathrm{MoS_2/vac}} - E^\mathrm{VBM}_{b-\mathrm{MoO_3/vac}} \nonumber \\
& \approx (\overline{E}^\mathrm{VBM}_{l-\mathrm{MoS_2}} - e\overline{V}_{\mathrm{H,vac}/l-\mathrm{MoS_2}}) - (\overline{E}^\mathrm{VBM}_{b-\mathrm{MoO_3}} - e\overline{V}_{\mathrm{H,vac}/l-\mathrm{MoO_3}})
\end{align}

% ---------- SI --------------------
%E^\mathrm{VBM}_{l-\mathrm{MoS_2/vac}} & = ``E^\mathrm{VBM}_{l-\mathrm{MoS_2}}" - ``E_{\mathrm{vac}/l-\mathrm{MoS_2}}" \nonumber \\
%& = ``E^\mathrm{VBM}_{l-\mathrm{MoS_2}}" - e\overline{V}_{\mathrm{H},l-\mathrm{MoS_2}} + e\overline{V}_{\mathrm{H},l-\mathrm{MoS_2}} - ``E_{\mathrm{vac}/l-\mathrm{MoS_2}}" \nonumber \\
%& = \overline{E}^\mathrm{VBM}_{l-\mathrm{MoS_2}} - e\overline{V}_{\mathrm{H,vac}/l-\mathrm{MoS_2}} \nonumber \\
%E^\mathrm{VBM}_{b-\mathrm{MoO_3/vac}} & = \overline{E}^\mathrm{VBM}_{b-\mathrm{MoO_3}} + e\overline{V}_{\mathrm{H},b-\mathrm{MoO_3}} - ``E_{\mathrm{vac}/b-\mathrm{MoO_3}}" \nonumber \\
%& = \overline{E}^\mathrm{VBM}_{b-\mathrm{MoO_3}}  - e\overline{V}_{\mathrm{H,vac}/b-\mathrm{MoO_3}} \nonumber
%& \approx \overline{E}^\mathrm{VBM}_{b-\mathrm{MoO_3}}  - e\overline{V}_{\mathrm{H,vac}/l-\mathrm{MoO_3}} \nonumber
% ------------------------------------
where the quantities with bar on top indicates referencing with respect to the macroscopic average either for slabs ($l-$) or for the bulk ($b-$), and $V_{\mathrm{H,vac}/A}$ indicates planar averaged Hartree potential in the vacuum region for slab $A$. Note that for the explicit $\mathrm{MoS_2}$ slab calculation, it does not matter whether the averaging of the Hartree potential is performed for the entire supercell or just the slab region. The key aspect in equation 1 is to identify $\overline{V}_{\mathrm{H,vac}/l-\mathrm{MoO_3}}$ with $\overline{V}_{\mathrm{H,vac}/b-\mathrm{MoO_3}}$ for large enough supercell.
The estimation of band offset in explicit heterostructures is much simpler and does not require any additional alignment. The band edges can be determined directly via Partial Density of States (PDOS) calculations. The heterostructures MoO$_3$/MoS$_2$ , MoO$_3$/MoTe$_2$ and CuSbS$_2$/CdS have the following lateral dimensions: 15.849(Å) x 11.091(Å), 11.887 (Å) x 3.697 (Å) and  7.603 (Å) x 12.146 (Å), respectively. Details of the choice of these dimensions have been provided in the SI. 
% ====================================================================================================  
\section{Results and Discussion}
\subsection{Band alignment in MoS$_2$/MoO$_3$ heterostructure}
We investigate the combined band structure of MoS$_2$ and MoO$_3$ by: (1) aligning the band edges of the individual materials with reference to the vacuum level (2) constructing a supercell with a few layers of MoS$_2$/MoO$_3$ heterostructure.
In the case of the isolated systems, we consider a bi-layer of MoS$_2$ and bulk MoO$_3$. The details of the band alignment procedure are provided in the method section. Essentially, the two key components for aligning bands between isolated systems are (1) band gap, $E_g$ and (2) VBO. It turns out that the choice of the DFT functional can have a significant effect on both the factors. Hybrid functionals typically provide good estimates for $E_g$s \cite{brothers2008accurate,henderson2011accurate,xiao2011accurate} and in some cases, provide more accurate VBOs too. \cite{weston2018accurate,huang2019band} However, given the computational cost, it is impractical to compute the full macroscopic averages relative to the vacuum with hybrid functional since one needs to consider a relatively large supercell (Figure S1). To circumvent this issue, Weston \textit{et al.} proposed computing the $E_g$s and VBM edges using HSE06 functional, while estimating the computationally demanding macroscopic averages with the PBE functional. \cite{weston2018accurate} Alternatively, the VBM edges and macroscopic averages can be estimated with GGA (PBE) functional while the band gaps are obtained from experimental data \cite{weston2018accurate,moses2011hybrid}. We compare these combinations schematically for bi-layer (2\textit{l}) MoS$_2$/bulk (\textit{b}) MoO$_3$ heterojunction in Figure-1a,b,c. In Figure 1d, we combine the VBM edges (with respect to vacuum) computed with PBE functional with the band gaps estimated with a hybrid functional, HSE06. As can be seen clearly, the type-II behaviour is realized (Figure 1c,d) when the VBO is less than the $E_g$ of MoO$_3$, the material with a lower VBM, when referenced to vacuum. Previous calculations suggest that even when the absolute band edges are computed with HSE06 functional for several isolated layers, one still ends up with a type-III arrangement, analogous to Figure-1b, instead of the type-II.\cite{shahrokhi20212d} Hence, one should treat the type-II alignments obtained in Figure-1c,d as constructs, only to reproduce the experimental observations, rather than as predictive tools. \\

\begin{figure}
    \centering
    \includegraphics[width=4.27in]{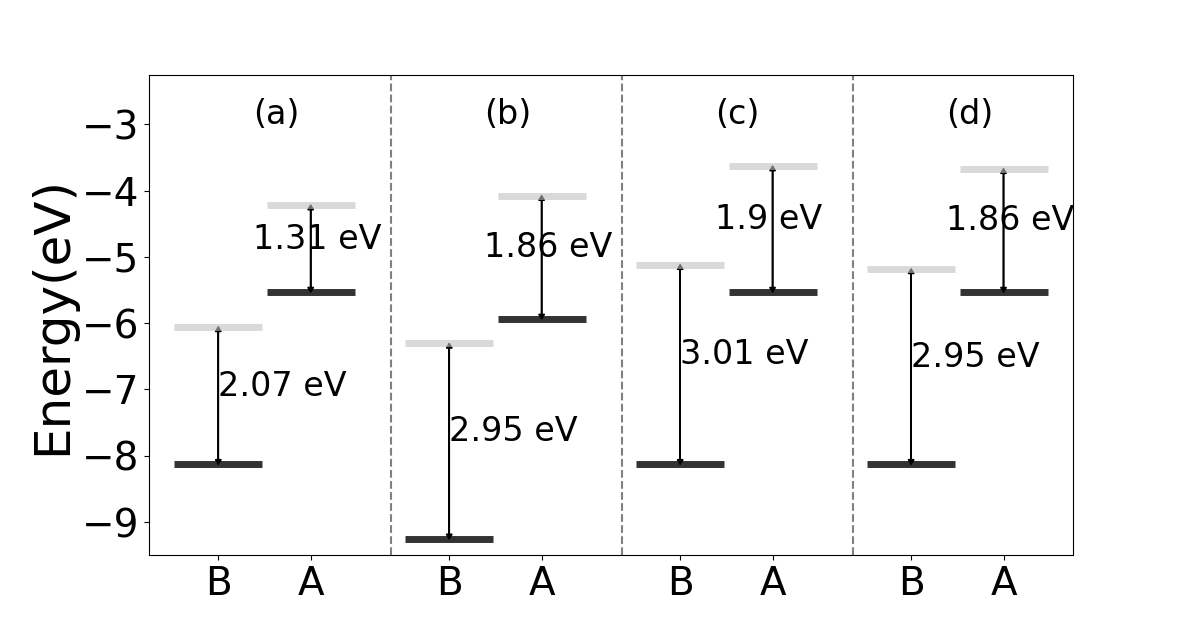}
    \caption*{\textbf{Figure 1}. Estimation of band alignment between 2$l$-MoS$_2$ (A) and $b$-MoO$_3$ (B) with different protocols to compute the VBOs and $E_g$s. (a) Both VBOs and $E_g$s are computed with PBE (b) VBMs and $E_g$s are computed with HSE06 but the macroscopic averaging employing slabs is performed with PBE (c) VBOs are computed with PBE but $E_g$s are obtained from previously reported experimental data on isolated layered MoS$_2$ \cite{ganatra2014few} and bulk MoO$_3$ \cite{chen2010single} (d) VBOs are computed with PBE but $E_g$s are estimated with HSE06. }
\end{figure}

In order to understand processes that require the physical presence of a certain type of heterojunction, one cannot treat the two materials separately. In those cases, it will be highly useful if the proper band alignment is present in the supercell with the explicit heterostructure, where the electronic structure is computed with either GGA or hybrid functional. The use of a hybrid functional does not always guarantee that a type-II arrangement will be obtained. For the MoS$_2$/MoO$_3$ explicit heterojunction, akin to the isolated case, one ends up with a type-III arrangement irrespective of whether PBE \cite{kc2016electronic} or HSE06 \cite{shahrokhi20212d} functional is employed. Moreover, a supercell for the explicit heterostructure might require a large lateral dimension to reduce the lattice strain between the two materials and hence, a GGA functional will be significantly more cost-effective.  \\

\subsection{The unconventional layering scheme}
In this study, we show that the type-II alignment can be realized in both the isolated and the explicit heterojunctions even with PBE functional if an alternative layer slicing scheme, referred to henceforth as the unconventional layering (uL) scheme, is employed for MoO$_3$ (illustrated in Figure 2b). This approach considers more distantly spaced layers as a single unit, leading to a slab with higher surface energy than the conventional one. The unit cell structures of MoO$_3$ for the two different types of layering and the corresponding band alignments with an isolated 2\textit{l}-MoS$_2$, computed with PBE only, are shown in Figure-2. Other combinations, as noted in Figure-1 (Table-S1), are provided in Figure-S2 (Table-S2). Unlike the conventional cut, the uL-scheme shows type II arrangement for all the combinations. Comparing Figures 2c and 2d, one can see that the band gap remains same in both the cases, as it should. The uL layering primarily affects the macroscopic average relative to the vacuum, thus altering the VBO without changing the individual band structures.\\
\begin{figure} %[h!]  % h! to keep image at same posi as latex 
    \centering
    \includegraphics[width=4.27in]{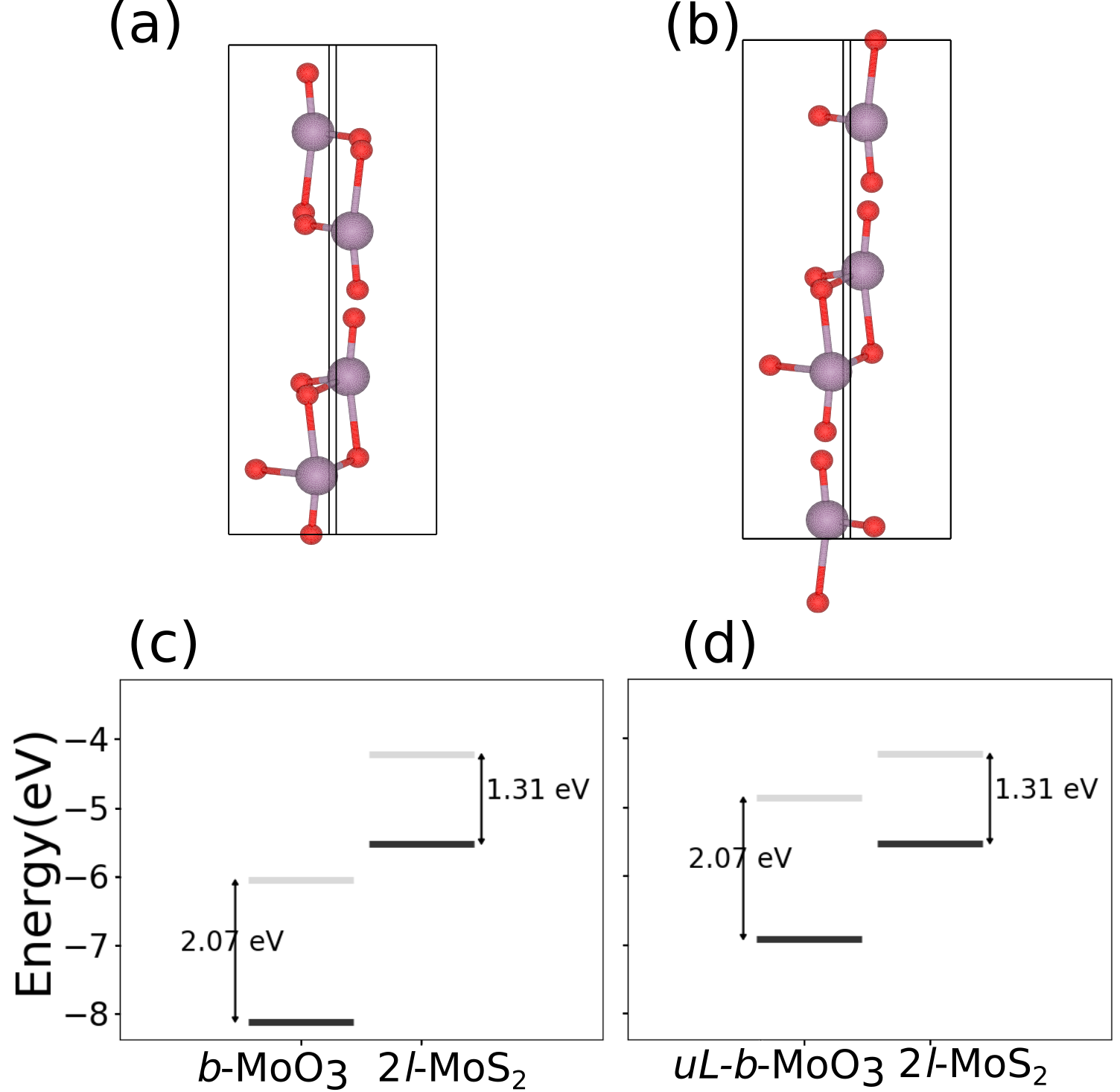}
    \caption*{\textbf{Figure 2}. Unit cell for (a) conventional and (b) unconventional(uL) layering. The corresponding band offsets, computed with PBE, are shown in (c) and (d), respectively.} 
\end{figure}

In the case of the explicit heterojunction, the same lattice strain would be present irrespective of the layering scheme since the lateral dimensions remain the same. Given the results obtained in the case of the isolated case, one would hope that the uL-scheme might exhibit type-II alignment in the case of explicit heterojunction also. The supercells for explicit heterostructure between MoO$_3$ (conventional and uL-) and MoS$_2$ are shown in Figure 3 a,c, while the corresponding PDOS plots are drawn in Figure 3 b,d, respectively. It can be clearly seen that while the band alignment in the conventional layering is type-III, the uL-scheme generates the desired type-II arrangement. It is to be noted that this artificial layering scheme will not be applicable in modeling those experimental set-ups where direct chemical interaction with the MoO$_3$ surface cannot be ignored.\\

\begin{figure}
    \centering
    \includegraphics[width=4.27in]{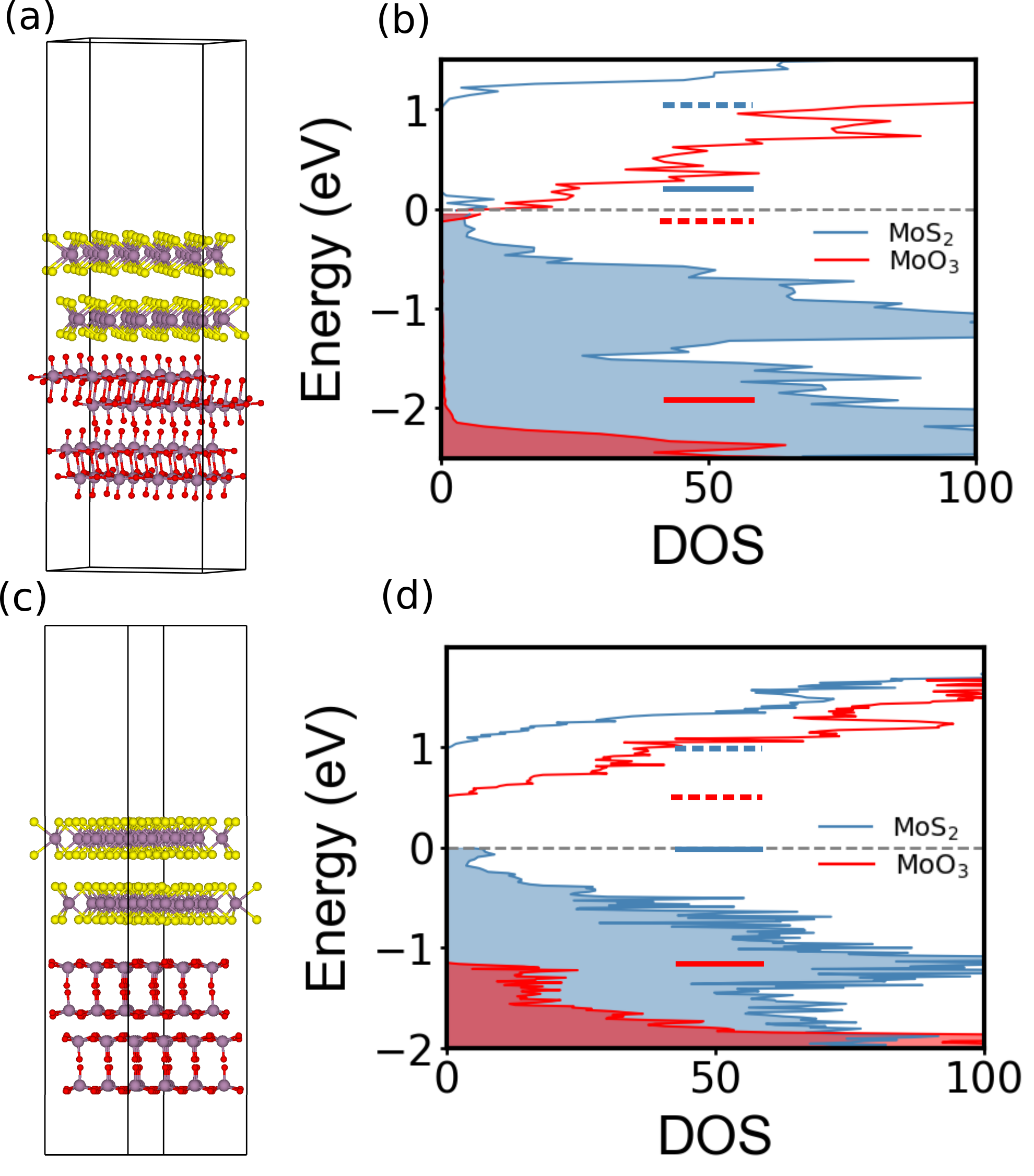}
    \caption*{\textbf{Figure 3}. (a) Supercell of 2\textit{l}-MoS\textsubscript{2}/MoO\textsubscript{3} and (c) 2\textit{l}-MoS\textsubscript{2}/ uL-MoO\textsubscript{3} heterostructures. Corresponding PDOS are shown in (b) and (d), respectively. The PDOS plots are resolved into bands on MoS$_2$(blue) and MoO$_3$(red). The dashed horizontal line marks the energy level of the highest occupied orbital, which is also set to `0' in the energy scale.  The band edges are identified by horizontal markers, solid for VBM and dashed for CBM.}
\end{figure}

\begin{figure}
    \centering
    \includegraphics[width=3.27in]{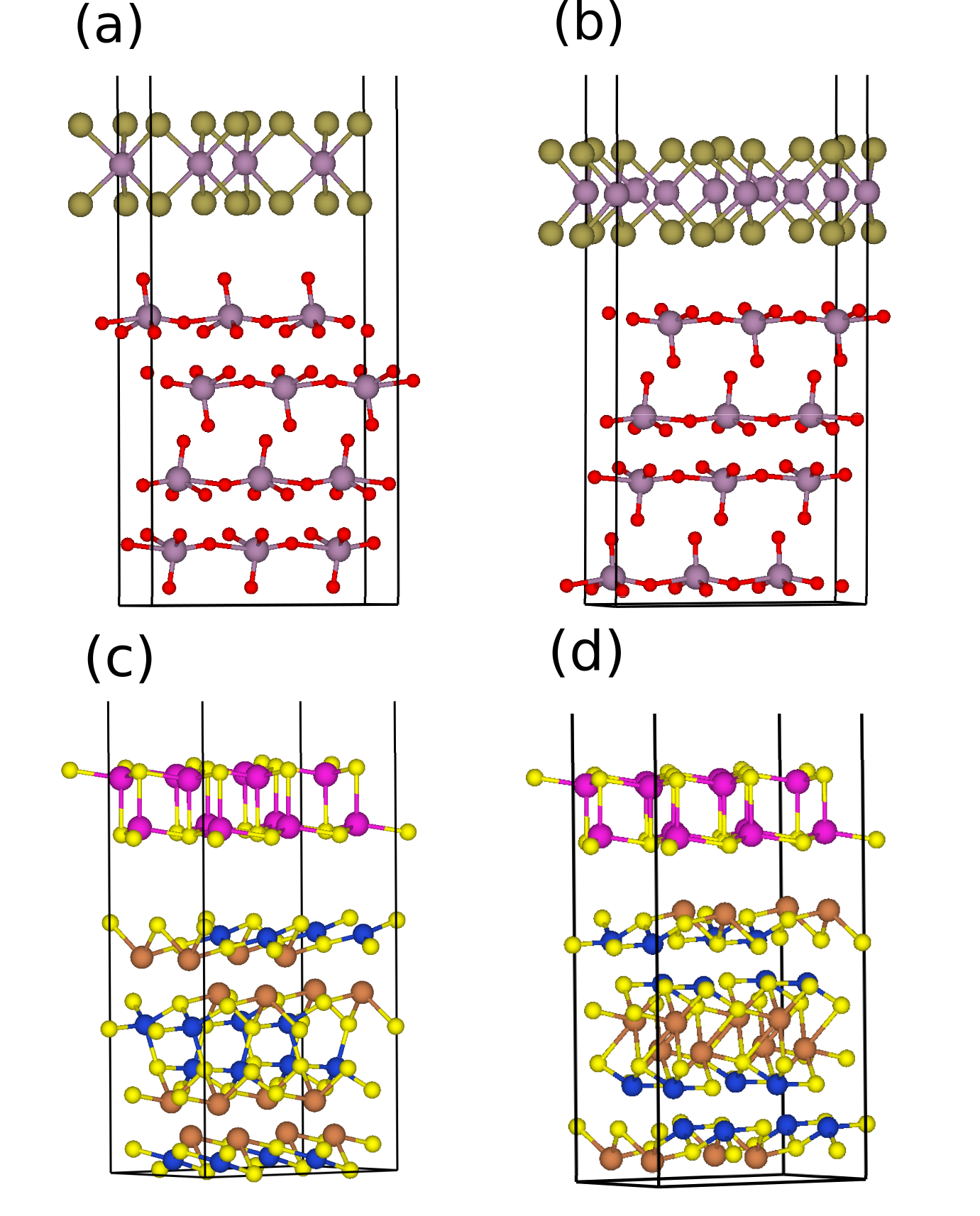}
    \caption*{\textbf{Figure 4}. Supercell structures of (a)  MoTe$_2$/MoO$_3$ (b) MoTe$_2$/uL-MoO$_3$ (c)  CdS/CuSbS\textsubscript{2}, and (d) CdS/uL-CuSbS\textsubscript{2}.}
\end{figure}

\subsection{Extension to other systems }
To show the general applicability of the artificial layering method, we explore different heterojunctions with 2D materials where experimental results are available. This list includes (1) MoO$_3$/MoTe$_2$,\cite{qi2024ultrafast} and (2) CdS/CuSbS$_2$\cite{wang2021cusbs,su2018earth} heterojunctions. The conventional and uL-schemes for MoO$_3$/MoTe$_2$ and CdS/CuSbS$_2$ heterostructures are shown in Figures 4a,b and c,d, respectively. The corresponding PDOS plots are shown in Figures 5 a,b and c,d, respectively. Structures and band shifts for other layered TMD materials, for which we could not identify any experimentally reported type-II partners, are provided in Figures S3 and S4. Similar to the MoS$_2$/MoO$_3$ case, the shifting of the bands for the uL-scheme can be explained by the higher surface energies of the layers. The layers with higher surface energies also have lower work functions, defined here as the difference in energy between the vacuum level and the VBM. A comparison of the work functions of the two types of layering schemes for MoO$_3$ and CuSbS$_2$ is shown graphically in Figure-6a and b, respectively. 
\begin{figure}
    \centering
    \includegraphics[width=4.27in]{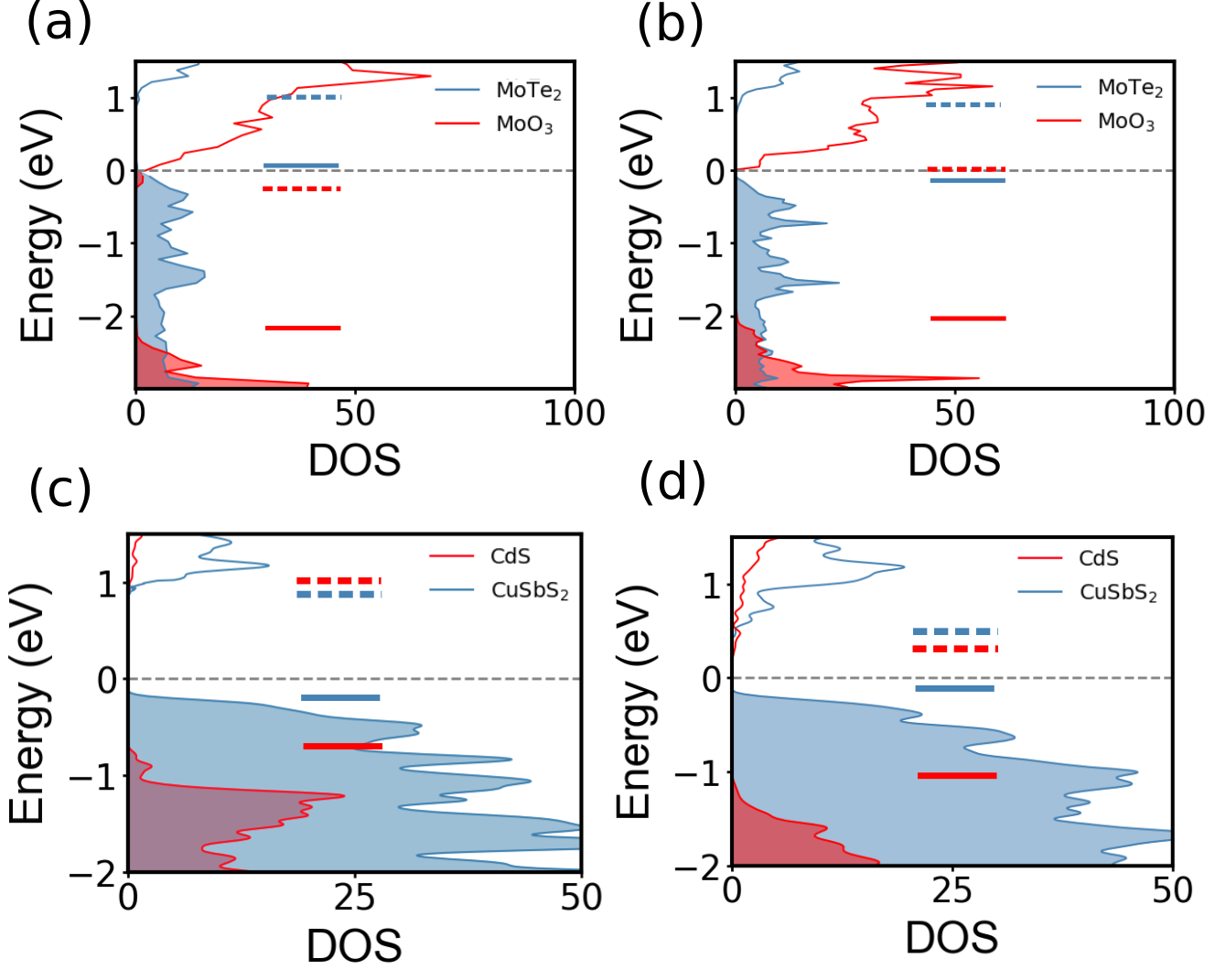}
    \caption*{\textbf{Figure 5}. DOS plots of (a) MoTe$_2$/MoO$_3$ (b) MoTe$_2$/uL-MoO$_3$ (c)CdS/CuSbS\textsubscript{2} (d) CdS/uL-CuSbS\textsubscript{2}. }
\end{figure}

\begin{figure}
    \centering
    \includegraphics[width=4.27in]{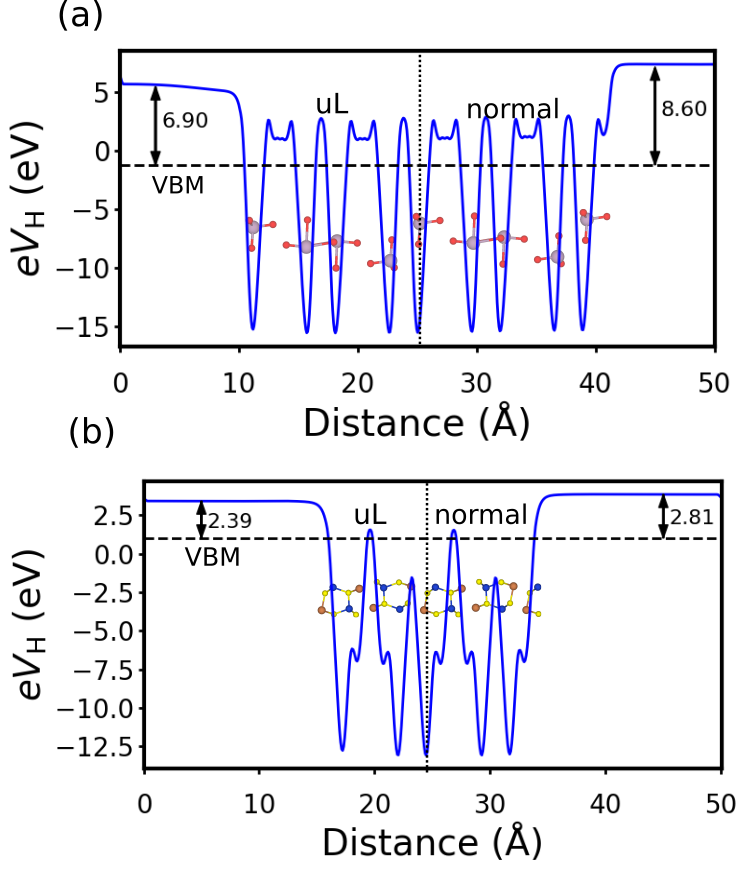}
    \caption*{\textbf{Figure 6}. 1D profiles of planar averaged Hartree potential for a combination of conventional and unconventional layers to compare their work functions in the case of (a) MoO$_3$  and (b) CuSbS\textsubscript{2}.}
\end{figure}

\section{Conclusions}
In this study, we developed a computational methodology to model the type-II band alignment between layered TMDs. This methodology relies on an alternative layering scheme with a lower work function that shifts the corresponding VBM level to higher values, thus changing the VBO between the two materials. We applied this methodology successfully for a few explicit heterostructures where experimental data is available. We believe that this computational scheme can be widely employed to study chemical reactions on type-II heterojunctions where such a layering scheme can be constructed.  

\begin{acknowledgement}

%Please use ``The authors thank \ldots'' rather than ``The authors would like to thank \ldots''. The author thanks Mats Dahlgren for version one of \textsf{achemso}, and Donald Arseneau for the code taken from \textsf{cite} to move citations after punctuation. Many users have provided feedback on the class, which is reflected in all of the different demonstrations shown in this document.
We acknowledge support of the Department of Atomic Energy, Government
of India, under Project Identification No. RTI 4007.

\end{acknowledgement}

\begin{suppinfo}

The SI contains supercell structure used for computing macroscopic average of $b$-MoO$_3$ (Figure S1), tables regarding band-offset data for 2$l$-MoS$_2$/$b$-MoO$_3$ (Table S1) and 2$l$-MoS$_2$/uL-$b$-MoO$_3$ (Table S2), schematic diagram of band alignment corresponding to isolated 2$l$-MoS$_2$/uL-$b$-MoO$_3$ heterostructure (Figure S2), bulk unit cells of normal and uL-layering for PbCN$_2$, CuSbS$_2$, and TeO$_2$ (Figure S3), their corresponding band positions (Figure S4), additional details regarding equation 1 and generation of the explicit heterostructures reported in the paper.

\end{suppinfo}

%\bibliography{achemso-demo}
\bibliography{RH_SA_SG_typeII_120824}
\end{document}